\documentclass[useAMS,usenatbib]{mn2e}
\usepackage{latexsym,graphicx,natbib}

\def\simless{{\th \rlap{\raise 0.5ex\hbox{$\scriptstyle  {<}$}}
    {\lower 0.3ex\hbox{$\scriptstyle  {\sim}$}} \th }}  
\def\simgreat{{\th \rlap{\raise 0.5ex\hbox{$\scriptstyle  {>}$}}
    {\lower 0.3ex\hbox{$\scriptstyle  {\sim}$}} \th }}  
\def\th{\thinspace}
\def\ts{{\raise 0.3ex\hbox{$\scriptstyle {\th \sim \th }$}}}
\def\be{\begin{equation}}
\def\ee{\end{equation}}
\newcommand{\ltaraw}{$\; \buildrel < \over \sim \;$}
\newcommand{\lta}{\lower.5ex\hbox{\ltaraw}}
\newcommand{\gtaraw}{$\; \buildrel > \over \sim \;$}
\newcommand{\gta}{\lower.5ex\hbox{\gtaraw}}

\newcommand{\kms}{{\rm\,km\,s^{-1}}}

\newcommand{\msun}{{\rm\,M_\odot}}

\def\ergs{{\rm\,erg\,s^{-1}}}
\def\mbh{M_{\rm BH}}
\def\vw{V_{\rm w}}
\def\vx{V_{\rm orb}}
\def\vr{V_{\rm rel}}
\def\gcm3{{\rm g\,\, cm^{-3}}}
\def\dotMw{\dot{M}_{\rm w}}

\def\arl{ a_{\rm {RL}}}
\def\abh{a_{\rm {BH}}}

\def\mdotbh{\dot{M}_{\rm{BH}}}
\def\amin{a_{\rm {min}}}
\def\amax{a_{\rm {max}}}

\def\kms{\rm {km\,s^{-1}}}
\def\au{\rm {AU}}

\newcommand{\apj}{ApJ}
\loadboldmathitalic 
\title[Radio pulsars around intermediate mass black holes]
{Radio pulsars around intermediate mass black holes
in super stellar clusters}
\author[A. Patruno, M. Colpi, A. Faulkner, A. Possenti]
{
A. Patruno$^{1,2}$\thanks{apatruno@science.uva.nl}, M.Colpi$^2$,
A. Faulkner$^3$ \& A. Possenti$^4$ \\
$^1$ Astronomical Institute ``A. Pannekoek'', University of Amsterdam,
Kruislaan 403, 1098 SJ, The Netherlands\\ 
$^2$Dipartimento di Fisica G. Occhialini, Universit\`a di Milano
Bicocca, Piazza della Scienza 3. I-20126 Milano, Italy\\
$^3$University of Manchester, Jodrell Bank Observatory, Macclesfield,
Cheshire \\ 
$^4$INAF, Osservatorio Astronomico di Cagliari, Poggio dei
Pini, Strada 54, Capoterra Italy 
}
\date{\today}
\begin{document} 
\maketitle 
\begin{abstract}
We study accretion in binaries hosting an intermediate mass black hole
(IMBH) of $\sim 1000 \msun,$ and a donor star more massive than
$15\msun$.  These systems experience an active X-ray phase
characterized by luminosities varying over a wide interval, from
$<10^{36}\ergs $ up to a few $10^{40}\ergs$ typical of the ultra
luminous X-ray sources (ULXs).  Roche lobe overflow on the zero-age
main sequence and donor masses above 20$\msun$ can maintain a
long-lived accretion phase at the level required to feed a ULX source.
In wide systems, wind transfer rates are magnified by the focusing
action of the IMBH yielding wind luminosities $\simgreat
10^{38}\ergs$.  These high mass-IMBH binaries can be identified as
progenitors of IMBH-radio pulsar (PSR) binaries.  We find that the
formation of an IMBH-PSR binary does not necessarely require the
transit through a ULX phase, but that a ULX can highlight a system
that will evolve into an IMBH-PSR, if the mass of the donor star is
constrained to lie within 15 to 30$\msun.$ We show that binary
evolution delivers the pre-exploding helium core in an orbit such that
after explosion, the neutron star has a very high probability to
remain bound to the IMBH, at distances of 1-10 AU.  The
detection of an IMBH-PSR binary in the Milky Way has suffered, so far, from
the same small number of statistics limit affecting the population of
ULXs in our Galaxy. Ongoing deeper surveys or next generation
radio telescopes like SKA will have an improved chance to unveil such
intriguing systems. Timing analysis of a pulsar orbiting around an
IMBH would weigh the black hole in the still uncharted interval of
mass around $1000\msun.$

\end{abstract}

\section{Introduction} 

Recent high resolution X-ray imaging and spectroscopic studies with
{\it Chandra} and {\it XMM},
have led to the discovery of a large sample of a new class of compact
 sources with luminosities in the
interval between $3\times 10^{39}\ergs$ and $10^{41}\ergs$, that are
in excess of the Eddington limit of a stellar-mass black hole
of 20$\msun$ (Fabbiano 1989; see Mushotzky 2004 for a critical review).  
These 
sources can find a simple interpretation in the hypothesis
that intermediate mass black holes (IMBHs) exist  
with mass $10^{2}\msun-10^{4}M_{\odot}$ 
accreting from a companion star in binary systems (Fabbiano 1989;
Mushotzky 2004; Miller \& Colbert 2004).  The detection of a cool-disc
thermal spectral component in a number
of ULXs (Miller et al. 2003; Zampieri et al. 2004;
Miller, Fabian \& Miller 2004; Kaaret et al. 2004; Cropper et
al. 2004; Dewangan et al. 2004), the properties
of the optical and radio counterparts (Koerding, Colbert \& Falcke
2005; Liu et al. 2004; Zampieri et al. 2004; Kaaret et al. 2004; Soria
et al. 2005; Miller, Mushotzky \& Neff 2005), and the timing behaviour
of at least one source in the starburst galaxy M82 (Strohmayer \&
Mushotzky 2003; Fiorito \& Titarchuk 2004), support this view.
The IMBH hypothesis however does not represent the only possibility to
explain the emission of ULXs, since mechanical beaming working in a
thick disc around a conventional stellar-mass black hole, or Doppler
boosting from a jet in a microblazar could produce the same range of
observed luminosities (King et al. 2001; Koerding et al. 2002;
Mushotzky 2004). Moreover, in a recent work of Rappaport, Podsiadlowski
\& Phfal (2005) the authors use binary evolution
calculations to show how the largest part of the ULX population may be
 explained with a stellar mass black hole emitting at a super
 Eddington rate of $\sim 10$ without the requirement of an IMBH 
(see also Podsiadlowski, Rappaport \& Han 2003 and Phfal,
Podsiadlowski \& Rappaport 2005 for an extended
study on the evolution of stellar mass black hole binaries). 
This is also in agreement with another study of King \& Dehnen
(2005) where the authors claim that the luminosities of a large sample
of ULXs could be explained using helium enriched matter and mechanical
beaming with a stellar mass black hole.  
But, all the alternatives to the IMBH hypothesis meet with strong difficulties
when the luminosity of a ULX is in excess of $\sim 10^{40}\ergs$,
since, in this case, beaming under rather extreme
conditions should be at work to match with the
observations, or super Eddington factors greater than $\sim 10$ could
be difficult to achieve.  
On the other hand, although the hypothesis of an IMBH
explains naturally many of the observational clues, it clashes with
the problem of providing a viable mechanism of formation.  Until now, two
possibilities have been proposed: the formation of an IMBH through
runaway collisions among massive stars undergoing fast dynamical
segregation, in the core of a dense super star cluster (see Porteigies
Zwart et al. 2004; Gurkan, Freitag \& Rasio 2004), or the
wandering of an IMBH, relic of a zero metallicity population III star (Abel et
al. 2000).  In the first case, the giant star that forms in the core
of the dense star cluster, collapses into an IMBH. 
This can occur in a starforming region, and the result is
based on very large detailed N-body simulations (Portegies Zwart et
al. 2004a). The case of wandering IMBHs relic of the early 
assembly of haloes in a currently starforming galaxy    
is uncertain, in particular the capture of gas or of a
star to ignite accretion (Volonteri \& Perna 2005).

In order to solve the controversy on the real existence of IMBHs in
ULXs, the only secure route would be the determination of the optical
mass function, similar to the procedure for the stellar-mass
black holes in the Milky Way (Orosz et al. 2004). This is difficult
however, since ULXs are distant sources hosted in external (starburst)
galaxies for which the optical identification of the companion is
troublesome, and even in the lucky circumstance of good identification
(see Zampieri et al. 2004; Kaaret et al. 2004; Soria et al. 2004; Liu
et al. 2004; Miller et al. 2004) the optical spectrum is too noisy
to allow a mass estimate of the black hole.

In this paper we propose an alternative way to discover and weigh a
IMBH: it uses the detection of a young radio pulsar around an IMBH.  If ULXs
are indeed accreting IMBHs in binaries, their evolution would end with
a radio pulsar around the IMBH, if the mass of the donor star is in
the range which avoids the formation of a stellar-mass black hole or
of a white dwarf.  The detection of a radio pulsar orbiting around an
IMBH would provide, through timing, an unambiguous measure of its
mass.

At present, no young radio pulsars have been seen orbiting a heavy
invisible companion, but in the near future the search of these
hypothetical IMBH-pulsar systems (IMBH-PSR hereon) can be extended to
nearby galaxies, with LOFAR (R\"ottgering 2003) and SKA (Cordes et
al. 2004).  An IMBH-PSR system would lead to the discovery of black
holes in the still uninvestigated interval of masses between
$100\msun-10^4\msun$.  Stellar-mass black holes around pulsars have
long been considered to be the ``holy grail'' of compact star
binaries. The systems studied here are even more intriguing, given their
importance in discovering a new unexplored mass range crucial for
cosmology (Madau \& Rees 2001).  The discovery of a pulsar orbiting
around a stellar-mass black hole is expected in the next years, on the
basis of theoretical considerations concerning their population, and on
the observability of the radio pulsar signal (Lipunov et al.  1994,
Sigurdsson 2003, Lipunov et al. 2005).  On similar lines, there is
also the hope of detecting, despite the large interstellar electron
densities, a radio pulsar orbiting around the massive black hole of
$\sim3\times 10^6\msun$ hosted in the core of
our Galaxy (Pfahl \& Loeb 2004).  Thus, the issue of pulsars around
black holes is becoming of paramount importance.

In this paper we will assume the hypothesis of the formation of IMBHs
in young dense star clusters, and will start our evolution
 study just after the formation/capture of a high mass star around the
 IMBH, which could be the progenitor of an IMBH-PSR system. 
 In this framework, Hopman et al. (2004) considered the possibility
 that a passing star
is tidally captured by the IMBH in a stable, close, not plunging
orbit. Mass transfer may initiate, after circularization, while the
star is on the main sequence or evolving away from it.  Dynamical
capture of a massive star by an IMBH is a further possibility, as
shown by Baumgardt et al. (2004). In an exchange interaction of a
binary star, the IMBH can acquire a companion, likely a massive star,
given that the IMBH forms in a mass-segregated environment,
where stellar encounters play an important role.

The observational appearance of binaries hosting an IMBH has only been
partly explored, and mainly in the context of ULXs.  Portegies Zwart
et al.  (2004b) studied the evolution of an IMBH of 1000$\msun$
accreting from a donor star of mass between 5 to $15\msun$ (see also
Kalogera et al. 2004 for another binary evolutionary study). Typically,
light donors ($\simless 5 \msun$) do not transfer mass at a
sufficiently sustained rate to produce sources as bright as ULXs; only
the high mass stars ($\simgreat 10\msun$) on the main sequence or
beyond can provide luminosities in excess of
$10^{40}\ergs$. Thus,  IMBHs cannot easily be identified on the basis of
their X-ray activity: they can display a rather wide range of
luminosities, depending on the mass of the donor star, and on the mass
transfer mechanism, as it will be shown in this paper.  Only when they
outshine as bright ULXs can they become visible over the stellar-mass
black holes.

In this paper we address a number of issues, in the hypothesis  
that an X-ray and/or a ULX phase preceeds 
the one in which
the system appears as an IMBH-PSR binary:
\begin{itemize}
 \item  
What are the characteristics of the X-ray phase and of mass transfer
when considering massive donors around an IMBH?
\item Is the formation of a radio pulsar likely?
\item What is the probability that an asymmetric kick, imparted to
 the neutron star at birth, will unbind the system?
\item On which orbits the radio pulsar is released after the supernova
explosion? Can Doppler effects hamper the detection of the radio
signal?
\end{itemize}
In order to address these questions, we explore in $\S 2$ wind fed
accretion (WFA), and accretion through Roche lobe overflow (ROLF),
when considering donor stars with masses in excess of $15\msun.$ We infer
the typical luminosities of these high-mass IMBH binaries and the fate
of the mass tranferring star.  In $\S 3$ we explore natal kicks in
such exotic massive binaries, and determine probabilities of survival
of the binary. Taking into account the effect of the natal kick, we
also compute the typical eccentricity and semimajor axis of the
neutron star orbit, after the supernova
explosion in the binary.
In $\S 4$ we simulate the observability of an
IMBH-PSR in the Galaxy, considering the parameters of the recent
survey for radiopulsars in the Galactic disc using the 
Parkes radio telescope.  In $\S5$ we discuss our results and outline our
conclusions.

\section {Accretion onto IMBHs}

Here we explore the evolution of binaries with an IMBH of 1000 $\msun$
and stars having mass $M$ heavier than 15$\msun,$ so that we can be
confident that, despite mass loss, the donor star does not evolve into a white
dwarf. The helium core mass necessary to form a neutron star is in the
range $2.8M_{\odot}<M_{\rm He}<8M_{\odot}$, although some
uncertainites exist in these upper and lower limits (Tauris \& van den
Heuvel 2004).  In each evolutionary sequence, obtained by varying the
mass and the initial orbital separation, the star is evolved until
carbon ignition.  The mass transfer episodes are either driven by RLOF
during the main sequence phase; by RLOF during the 
rapid expansion of the star
when ascending the giant branch; or by WFA when the system is
detached.  This third possibility has always been overlooked in
previous studies, though it seems relevant to understand the different
phases of an accreting IMBH.

The evolution equations for the binary systems are solved using an
updated version of the Eggleton code (Eggleton 1971; Pols et al. 1995)
with a modified mass transfer rate for the WFA case, as explained
later in the text.  All the stars considered have Population I
chemical composition (Y=0.28, Z=0.02), radii $R$ and wind losses
$\dotMw$ obtained according to de Jager et al. (1988). The mixing
length parameter is set to $\alpha=2.0$, and the overshooting
constant $\delta_{ov}=1.2$ (Pols et al. 1998).  Using general
arguments, we then focus attention on the remnant left in the
evolution to study the detectability of a young neutron star revolving
around an IMBH.

\subsection {Wind Fed Accretion}

Consider a donor main sequence star of mass $M$ above $15\msun,$
moving on a circular orbit around an IMBH of mass $\mbh.$ Let $a$ be
the relative separation,
\be 
\vx \simeq 950\Bigg(\frac{\mbh+M}{1015\msun}\Bigg)^{1/2}
\Big(\frac{a}{1\au}\Big)^{-1/2}\rm\,km\,s^{-1}
\ee 
the relative orbital velocity, and 
\be
\vw \simeq 1070\Bigg(\frac{M}{15\msun}\Bigg)^{1/2}\Bigg(
\frac{R}{5R_{\odot}}\Bigg)^{-1/2}\rm\,km\,s^{-1}\label{eq.wind}
\ee
the stellar wind velocity, set equal to the escape velocity from the
stellar surface.

WFA dominates when the star is underfilling its Roche lobe radius,
i.e., when the binary separation $a$ exceeds $\arl,$ defined as  
\be
\arl={R [0.6q^{2/3}+\ln(1+q^{1/3})]\over 0.49q^{2/3}} 
\ee
where $q=M/\mbh$.
 
In high mass X-ray binaries, the mass of the donor usually exceeds the
mass of the black hole, hence, the wind particles mainly accrete onto the
black hole from unbound orbits, since $\vx\ll \vw$. Capture occurs
within a cylinder of radius comparable to $\abh\sim
{2G\mbh/\vw^2}\approx (\mbh/M)\,R.$ In the fluid approximation, this
leads to an accretion rate onto the hole of
 \be
\mdotbh^{\rm SL}\sim {1/ 4}\left ({a_{\rm BH}/ a}\right )^2 ({\vr/ \vw})\dotMw 
\ee 
(Shapiro \& Lightman 1976, Shapiro \& Teukolsky 1983), where
$\vr$ is the wind speed relative to the black hole
\be 
\vr^{2}=\vw^{2}+\vx^{2}+2\vx\vw \cos\alpha,
\ee  with 
$\alpha\in [0,\pi]$  the angle between the directions of $\vec{V}_{\rm
orb}$ and $\vec{V}_{\rm w}$. Equation (4) is 
valid when $a\gg \abh,$  and $\vx\ll\vw,$ so that $\vr\sim \vw.$

\begin{figure}
\centerline{
\leavevmode
\includegraphics[width=85mm]{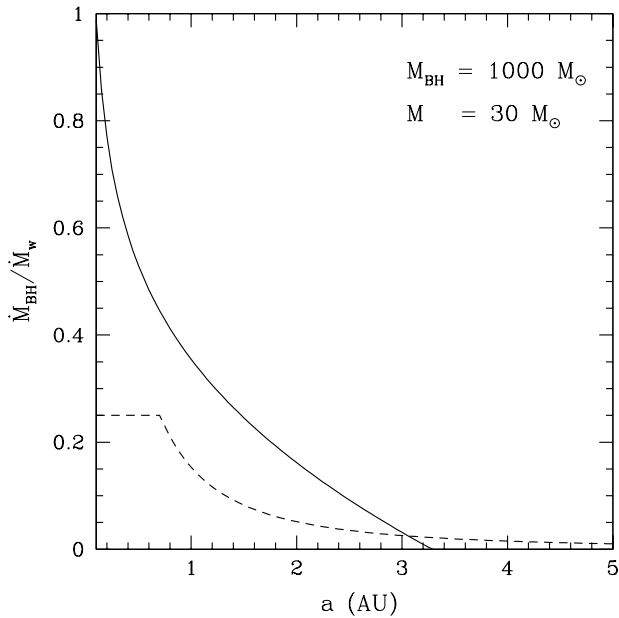}
}
\caption{$\mdotbh$ (solid line) and $\mdotbh^{\rm SL}$ (dashed line)
versus $a$ for a donor of 30$\msun$ and an IMBH of 1000$\msun$.  The
mass transfer rates are in units of the wind mass loss rate.  The
accretion rate onto the IMBH is the highest value between the two.
The figure is explained in the text.
}
\label{fig:f1}
\end{figure}

In a binary with an IMBH, we may have $\abh \simgreat a$ (or
equivalently $\vx\simgreat \vw$), i.e., the particles released from
the donor star surface can have kinetic energies such to be bound to
the IMBH.  This effect may lead to an accretion rate $\mdotbh$
comparable to the wind mass loss rate $\dotMw.$ In particular, if all
the wind particles are deep in the potential well of the IMBH, their
total energy $K+U=\vr^2/2-G\mbh/a$ is negative.  This implies a full
wind gravitational focussing by the IMBH, and an X-ray luminosity at
its highest value.  The limiting orbital separation for this to occur
corresponds to the case $K+U=0$ for the most energetic wind particles,
i.e., the ones having $\alpha=0.$ This yields $\vx=[1+\sqrt{2}]\vw,$
and accordingly, a binary separation
\be
a_{\rm min}= {G\mbh \over {(1+\sqrt {2})^2\vw^2} }\simeq 0.1\frac{\mbh}{M}R\ee
Below $\amin$ the wind is entirely accreted. When $\alpha=\pi$, the
condition $K+U=0$ corresponds to those particles having the minimum
relative speed to remain bound to the IMBH. This defines a critical
orbital velocity $\vx=[\sqrt{2}-1]\vw$ and in turn a limiting orbital
separation
\be 
a_{\rm max}\simeq {G\mbh \over {(\sqrt {2}-1)^2\vw^2}}\simeq
3{\mbh\over M}R
\ee
above which all wind particles are unbound.  In the interval $a_{\rm
min}<a<a_{\rm max}$ wind particles with $K+U>0$ escape the
gravitational field of the IMBH and this defines a loss cone of solid
angle $\Omega=2\pi(1-\cos\hat{\alpha}),$ with
$\hat{\alpha}=\cos^{-1}[(\vx^2-\vw^2)/(2\vw\vx)].$ Under these
conditions, the accretion rate onto the IMBH can be approximated as
\be
\mdotbh\sim {\rm{Max}}\left [ \left (1-{\Omega\over 4\pi}\right
)\dotMw; \mdotbh^{\rm SL} \right ]\label{mdotbh}
\ee
For the mass transfer rate of eq. (\ref{mdotbh}) even a very high mass
star can provide the ULX luminosities during the main sequence,
without the requirement that the donor is crossing the supergiant
phase.  In Figure~\ref{fig:f1} we illustrate this effect, plotting
accretion rates $\mdotbh$ and $\mdotbh^{\rm SL}$ for an IMBH of
1000$\msun$ and a donor star of 30$\msun$ during the main sequence.
The probability to observe this system is strongly increased due to
the larger time spent by the star on the main sequence.

In Figure 2, we show $\amin$ and $\amax$ as a function of the mass of
the IMBH, compared to $\arl,$ i.e., the distance at which the donor
star overfills its Roche lobe.  The light gray area corresponds to
orbital separations where WFA is in the Shapiro-Lightman limit, while
in the middle gray area, the wind is enhanced by the action of gravitational
focussing by the IMBH, and in the heavy gray area, it is fully captured, 
leading to a conservative system.  The black area is the RLOF zone.

\begin{figure}
\centerline{
\leavevmode
\includegraphics[width=85mm]{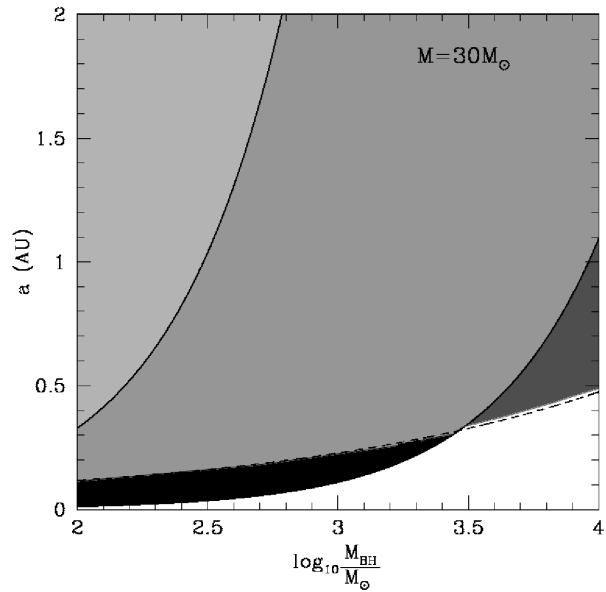} 
}
\caption{$\arl$ (dashed line), $a_{\rm min}$(lower solid line) and
$a_{\rm max}$ (upper solid line) versus $\mbh$ for a donor of
30$\msun$ on the ZAMS. WFA prevails above the dashed line.  If the
donor star is inside the gray shaded area, WFA is enhanced by the
gravity of the IMBH.  Black area ($\amin>\arl$) corresponds to RLOF,
while the heavy gray area is for accretion of the entire wind lost by
the donor star. Mild gray refers to a partial focussing of the
wind, while the light gray is the Shapiro-Lightman regime.  The
condition $a_{\rm min}=\arl$ 
defines a minimum IMBH mass for which WFA occurs under full
capture of the stellar wind, giving maximum WFA luminosities.  For a
30$\msun$ donor this minimum IMBH mass corresponds to $\sim
3000\msun$.
}
\label{fig:f2}
\end{figure}

In our WFA model, the possible formation of a circumbinary disc around
the donor star might change the quantity of wind accreted, although
the order of magnitude remains unchanged (see van den Heuvel 1994). 
We have followed the evolution of the binary under WFA, starting from
a separation of 1 AU, imposing angular momentum and mass loss from the
fraction of the wind escaping from the binary. The star is non
rotating and the wind leaves the donor isotropically, carrying away, in
the centre mass reference frame, a 
specific angular momentum $\dot{J}=h(\dot{M}_{\rm w}-\dot{M}_{\rm BH})$
with $h$ the specific angular momentum of the orbit (see Soberman,
Phinney \& Van den Heuvel 1997 for a detailed discussion).
We have computed the accretion rate onto the IMBH using eq. (\ref{mdotbh})
and a WFA luminosity adopting an efficiency factor of $\sim 0.1$ 
characteristic of disc accretion, since the captured wind has high
enough angular momentum to form a disc.

\begin{figure}
\centerline{
\leavevmode
\includegraphics[width=85mm]{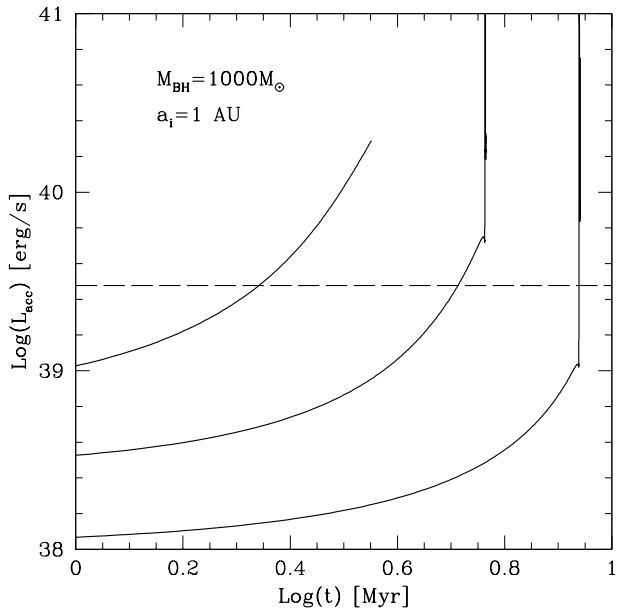}
}
\caption{From the top to the bottom lines: WFA luminosity against time
for donors with masses of 50$\msun$, 30$\msun,$ and 20$\msun$
respectively, and a black hole's mass of 1000$\msun.$ The initial
orbital separation is 1 AU. The dashed line is the limit to obtain an
ULX. The mass transfer begin on the ZAMS and the integration is
stopped when the star reaches the giant branch. At this point a RLOF
phase starts, producing the sharp spikes in the plot, with the
exception of the donor of $50M_{\odot}$ for which the code is stopped at
the end of the main sequence.  
}
\label{fig:f3}
\end{figure}

The run of the WFA luminosity as a function of time is shown in
Figure~\ref{fig:f3} for an IMBH of 1000$\msun$ in a binary with a
20-30-50$\msun$ star, respectively.  The luminosity is enhanced
relative to the value that would be predicted from a stellar-mass
black hole under equivalent conditions.  With time the WFA luminosity
rises mainly because of the instrinsic increase of the wind loss rate
when the star is transiting from the zero age
to the terminal age main sequence.  Only very massive stars can
provide mean luminosities in excess of $3\times 10^{39}\ergs,$ the
threshold defining ULX sources.

\subsection {Accretion through Roche lobe overflow}

Accretion via RLOF occurs during the main sequence when the star is
close enough to the IMBH to fill its Jacobi surface, or during the
rapid expansion of the star when ascending the giant or supergiant
branch.  In this section we explore the evolution for the former case,
i.e., for separations ranging between 0.18-0.25 AU for a 1000$\msun$
IMBH, and 15-50$\msun$ companion stars.

\begin{figure}
\centerline{
\leavevmode
\includegraphics[width=85mm]{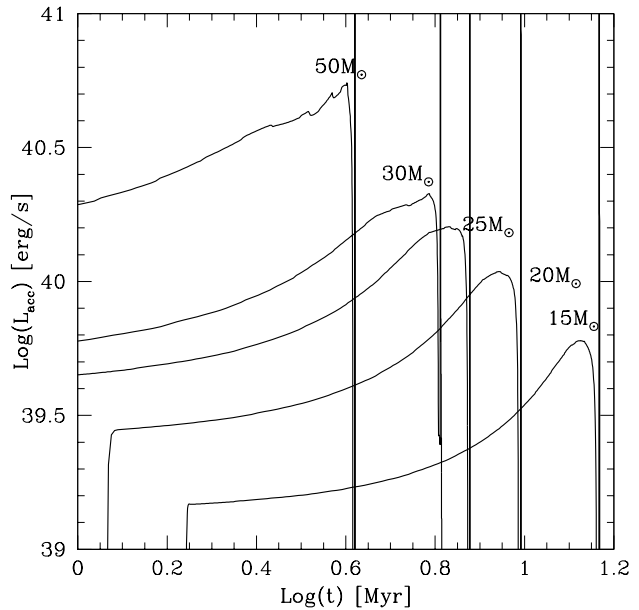}
}
\caption{Luminosity against time, for donors with masses of
$15,20,25,30,50M_{\odot},$ respectively, and for an IMBH of
$1000M_{\odot}$. The mass transfer begins near the zero age main
sequence and the integration is stopped before carbon ignition.
Stellar masses in excess of $\simeq20M_{\odot}$ can produce
luminosities above $10^{40} \ergs$ during the main sequence.
All  binaries appear to be persistent sources.
The dashed line represents the ULX limit.
Note that there is only one spike for each donor, corresponding to the
giant phase. The envelope of the donor is almost completely depleted during
this phase and the re-expansion after the helium burning is too small
to produce another contact phase.     
}
\label{fig:f4}
\end{figure}

Roche lobe contact remains stable along the entire evolution and gives
luminosities above $10^{39.5} \rm erg\,s^{-1}$ when the donor star has
a mass $M\simgreat 15\msun.$ Luminosities of $10^{40}\ergs$, typical of
the brightest ULXs, need a star at least as massive as 20$\msun,$ as
illustrated in Figure~\ref{fig:f4}.  Adopting the Dubus criterion for
assessing the stability of the disc against the thermal ionization
instability (Dubus et al. 1999), we find that RLOF onto an IMBH is
stable, under our conditions.  This ensures that high mass IMBH
binaries can remain bright for a relativley long time, between a few
million to ten million years.  This is opposite to the case of IMBH
binaries with donor stars having masses $\simless 10\msun$ as pointed
out by Portegies Zwart et al. (2004b), for which the luminosities are
transient and the X-ray phases last for a longer
time, comparable with
the donor lifetime.  Within the limits of the available time span
coverage of ULX observations, the apparent persistency in their
emission may be an indication that donor stars have masses in the
range considered in this paper.

\begin{figure}
\centerline{
\leavevmode
\includegraphics[width=85mm]{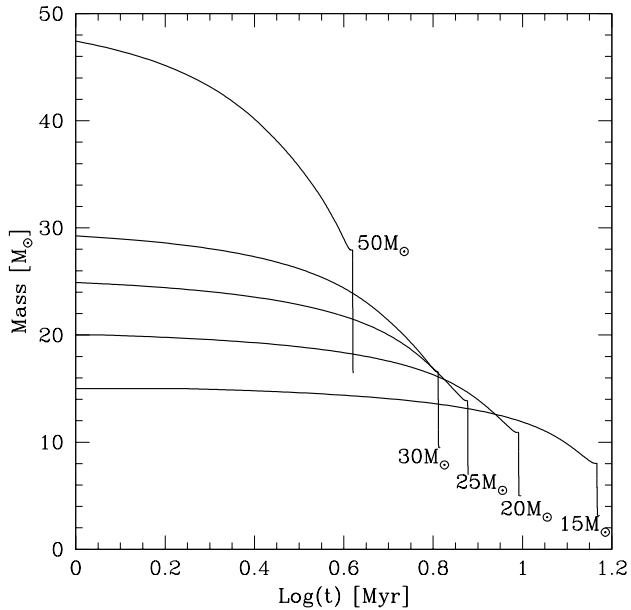}
}
\caption{Donor mass versus time during the whole life of the star
for RLOF systems. The mass loss is very rapid for a high mass star and
decrease mildly for a lighter star.  The plot shows binaries with donor
star masses of $15,20,25,30$ and $50M_{\odot}$ orbiting around a
$1000M_{\odot}$ IMBH. The very rapid short decrease near the end of
the curves is due to RLOF on the giant phase.
 Note that the initial mass corresponds to an
age of 1 Myr as we are using logarithmic scale on the time axis.
}
\label{fig:f5}
\end{figure}

The high mass transfer rates that establish during RLOF
dramatically change the mass of the star. Figure~\ref{fig:f5} shows the donor
 mass as a function of time. At the end of the stellar evolution, the
 core mass is almost equal to the mass of the whole star,
 since the mass transfer strips almost completely the entire stellar envelope
 during the RLOF phase. 
Donors with mass between 15$\msun$ and $\sim 25M_{\odot}$ on the zero 
age main sequence likely end their
lives with a core mass between $3.3$ and $7\msun$, probably producing
a neutron star.  The case of a $30M_{\odot}$ could be considered the
mass limit between the neutron star and the black hole formation,
since its final core mass is $\sim 9\msun$.  The fate of the heaviest
star here considered ($M\simgreat 30\msun$) is to collapse into a
stellar-mass black hole.

If RLOF is the mechanism leading to a ULX during the main
sequence, the donor star is expected to lie at an initial distance of
$a\simless 1$ AU from the IMBH.  After leaving the main sequence, the
star ascends the giant branch, while keeping on overflowing its Roche
lobe.  This generate the spikes in Figure~\ref{fig:f4} giving very
high luminosities.  The typical values of $a$ reached at the end of
the evolution are in the range of $\sim 2-5 $ AU.

If the initial orbital separation of a donor star is $a\simgreat 1$
AU, RLOF is avoided on the main sequence phase. The star, in exiting
the terminal age main sequence, may later overfill its Roche lobe due
to the sizable increase of its radius when ascending the giant branch,
initiating a phase of accretion. Mass transfer and angular momentum
conservation
 will drive the star further away from the IMBH, up to typical distances of
$8-10$ AU.

\section {Neutron star kicks and the survival of the IMBH-PSR binary}

In the RLOF case, the continuous increase of $a,$ implied by
conservative mass transfer, would bring the donor to distances of
several astronomical units. We may ask whether the explosion and the
asymmetric kick that accompanies the formation of the neutron star can
unbind the system.  Given the high inertia of the IMBH, a neutron star
has a greater chance to remain bound than in the case of a lighter
black hole companion.

To keep a binary bound, we need a kick velocity  below 
the limit given by the equation (Brandt \& Podsiadlowski 1994, 
Willems, Kalogera \& Henninger 2004)
\be
V_{k,l} = \left[
  {{G \left( M_{\rm BH} + M \right)} \over a_0} \right]^{1/2}
  \! \left[ 1 + \left( \! 2\, {{M_{\rm BH} + M_{\rm NS}} \over
  {M_{\rm BH} + M}} \right)^{1/2} \right]  \label{vkmax}
\ee
where $M_{\rm NS}$ is the mass of the neutron star formed after the
supernova explosion, and $M$ and $a_0$ are the mass of the donor star
and binary separation, just prior the supernova explosion.  In the
presence of the IMBH, the total mass remains almost constant before and
after the explosion, and so the most important parameter is the
orbital separation $a_0$.  To obtain a conservative estimate, we use
our largest value of $a_0$ which is about $10$ AU.  The kick velocity
limit is around $700\, \kms$.

To calculate the probability of survival of the IMBH-PSR system, we
adopt an analytical expression for the distribution of the kick
magnitudes (Hansen \& Phinney 1997, HP distribution)
\be
p(V_{k})=\sqrt{\frac{2}{\pi}}
\frac{V_{k}^{2}}{\sigma^{3}}e^{-V_{k}^{2}/\sigma^{2}}, 
\ee
where $\sigma$ is the velocity dispersion of each of the
components of the kick velocity.

Although there is not full agreement on the form of the kick magnitude
distribution, it is rather well established that speeds larger than
$\sim 700\, \kms$ are attained only in extreme rare cases (see also
Lyne \& Lorimer 1994, Arzoumanian et al. 2002, Hobbs et al. 2005).
Therefore we are confident that our calculations are not strongly
affected by the particular choice of the distribution, since what is
interesting here is the existence of a cutoff at high velocities more
than the specific form of the distribution itself.

We use the following equation to calculate the probability of survival
of our system, obtained by integrating the HP distribution between zero
and our kick velocity limit:
\be
\frac{\int_0^{V_{k,l}} p(V_{k})dV_{k}}{\int_0^{\infty}
p(V_{k})dV_{k}} ={\rm Erf}\Big(\frac{V_{k,l}}{\sigma}\Big)-
\frac{2V_{k,l}}{\sqrt{\pi}\sigma}e^{-2V_{k,l}^{2}/\sigma^{2}}
\ee
Using a value for $V_{k,l}$ of $700\,\kms$ and a value for $\sigma$ of
152 $\kms$ (Hobbs et al. 2005), we infer a probability of survival
greater than 99\%.  Therefore after the supernova explosion we can be
confident that the system is still bound.

After the explosion the orbital parameters change.  To obtain the
distribution of post-supernova orbital separations and eccentricities,
we introduce, following Kalogera (1996), the three dimensionless
parameters that are necessary to constrain the properties of the
binary after the explosion:
\begin{eqnarray}
\alpha&=&a_f/a_0\\
\beta&=&\frac{\mbh+M_{\rm NS}}{\mbh+M}\\
\xi&=&\frac{\rm \sigma}{\vx}
\end{eqnarray}
where $a_0$ is the pre-explosion binary separation and $a_f$ its
post-explosion value, $\beta\sim 1$ the binary total mass ratio, after
and prior explosion.

\begin{figure}
\centerline{
\leavevmode
\includegraphics[width=90mm]{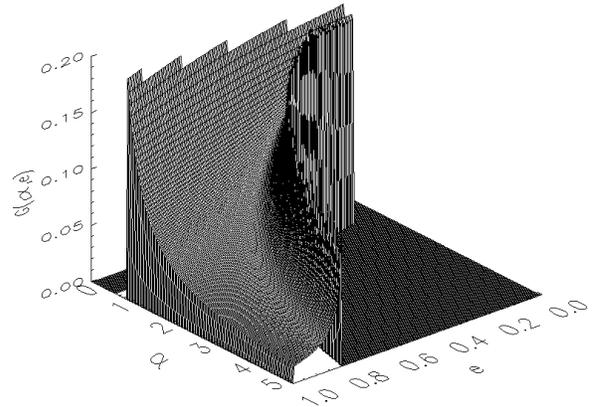}
}
\caption{Distribution of post-supernova systems over $\alpha$ and $e$,
normalized as in equation (14) of Kalogera (1996), calculated for
$\beta\simeq 1$ and $\xi=0.3$.
}
\label{fig:f6}
\end{figure}

\begin{figure}
\centerline{
\leavevmode
\includegraphics[width=85mm]{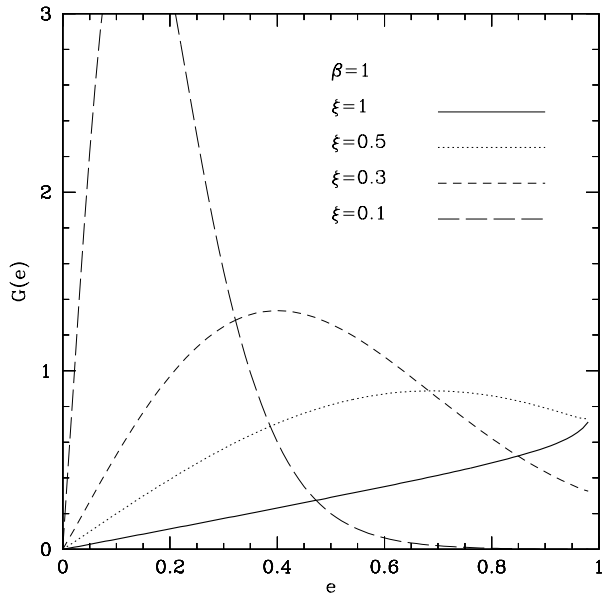}
}
\caption{Distribution of post-SN systems over eccentricites, calculated
for different values of $\xi$ and for $\beta\simeq 1$. For $\xi=1$ the
systems are highly eccentric, while they became less eccentric with
decreasing $\xi.$  
}
\label{fig:f7}
\end{figure}

\begin{figure}
\centerline{
\leavevmode
\includegraphics[width=85mm]{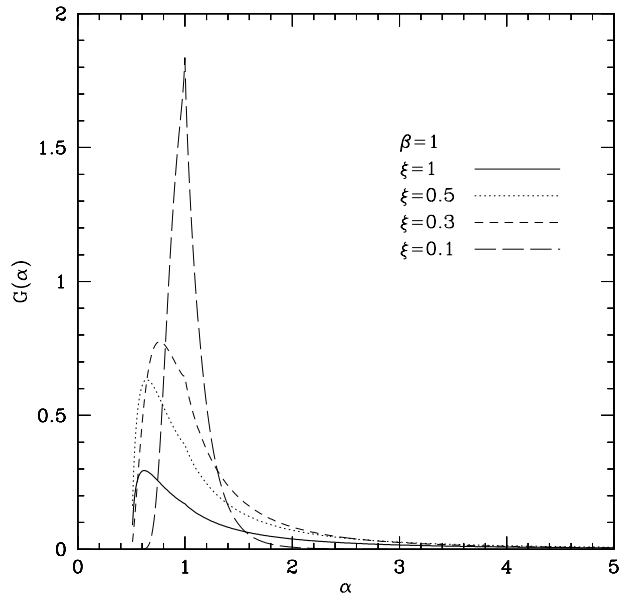}
}
\caption{Distribution of post-SN systems over dimensionless orbital
separation, calculated for different values of $\xi$ and for
$\beta\simeq 1$. For $\xi=1,$ the most probable post-SN semimajor axis
is about $0.5a_{0}$, while for $\xi<<1$ the most probable semimajor
axis is similar to the pre-SN orbital separation.
}
\label{fig:f8}
\end{figure}

Typical pulsar 1D speeds peak at $152 \kms$ (Hobbs et al. 2005).
Given the high relative orbital velocity (eq. 1) with respect to
$\sigma$, $\xi\simless 1 $ in our case.  Figure \ref{fig:f6} shows the
probability distribution over $\alpha$ and post-encounter eccentricity
$e$, computed for $\beta\sim1$ and $\xi=0.3$, while Figure
\ref{fig:f7}, and Figure \ref{fig:f8} give the distribution over $e$
(integrated over $\alpha$) and $\alpha$ (integrated over $e$)
respectively, for selected values of $\xi$.

Post-supernova binaries populate only a restricted area of the
$\alpha-e$ plane giving minimum acceptable post-supernova separations
$a_f=(1/2)a_0$ for $\xi\simeq 1$ and $a_f=a_0$ for $\xi<< 1$ according
to the fact that the post-supernova orbit must include the position of
the two stars just prior to explosion.  Because we have found in
Section 2.2 that the likely pre-supernova orbital separation falls in
the range 2-10 AU, final values of $a_f$ range between $\sim 1-10$ AU.

The small eccentricity of these systems is a direct consequence of the
very high mass of the IMBH, which produces higher orbital velocities
than in normal binaries, and a value of $\xi<1$ also for the largest
orbital separations.  Another consequence of the large mass of the
black hole is that the binary can be always defined as an hard one
(Heggie 1975) and thus the dynamical interactions in the star cluster
tend to decrease the orbital separation of the system, although there
are some uncertainities in the real effect of this process as stated
by Hopman \& Portegies Zwart (2005).

Neglecting this uncertain effect, the production of gravitational
waves seems to be unimportant, because the pulsar is revolving on a
mildly eccentric orbit at a large separation.  The in-spiral time for
emission of gravitational waves is extremely large: $t_{\rm GW}\simgreat
3\times 10^{10}$ yrs for $e=0.4$, and only for very high eccentric
systems ($e=0.9$) it is $t_{\rm GW}\sim 6\times10^{7}$ yrs.

The lifetime of a super star cluster is about $10^{8}$ yrs (see Hopman \&
Portegies Zwart 2005), which
is comparable to the lifetime of the binary even if the eccentricity
produced by dynamical interactions becomes $\sim 0.9$.  Thus we have a
very small probability of a merger in a super star cluster 
by gravitational waves
emission, and the observability in the radio band of our system is not
compromised by this effect.  This is in agreement with the results of
Hopman \& Portegies Zwart (2005) which found that with an IMBH of
$\sim 1000\msun$, only a fraction around $2\%$ would merge 
 in an Hubble
time.

\section{Simulating pulsars around IMBH}

The search of pulsars in binary systems is more difficult than
for isolated pulsars, because of changes in the
line-of-sight velocity due to orbital motion cause varying observed
pulse periods. Most of
the algorithms for searching for periodicities in a time series use
a Fast Fourier Transform (FFT) to produce a frequency
spectrum; a varying received pulsation frequency spreads the detection
over many spectral bins. This reduces the signal-to-noise ratio (s/n) of
a detection with respect to that of an isolated pulsar with the same
intrinsic flux and rotational period. 

We have investigated the consequencies of this effect on the
probability of detecting a pulsar orbiting an IMBH during the Parkes
Pulsar Multibeam survey (Manchester et al. 2001), i.e. the deepest and
most successful large scale survey for pulsars in the Galactic disc
completed so far, resulting in the discovery of more than 700 new
sources (Faulkner et al. 2004). We have generated many sets of
simulated time
series containing a pulsating signal (with a duty cycle of $5\%$, as
statistically appropriate for non-recycled pulsars) of a
pulsar of $1.4~\msun$ orbiting an IMBH of $1000\msun.$ The other
binary parameters have been chosen to span the most likely values of
the distribution plotted in Figure \ref{fig:f6}, i.e. orbital
separations in the range 1-10 AU (corresponding to orbital periods
from $\sim 10$ days to $\sim 1$ yr) and eccentricity from 0 to 1 in
steps of 0.05. The sampling time (0.25 ms) and the total number of
samples ($2^{23}$) are the same as the observations of the Parkes
Pulsar Multibeam survey. Spin periods have been chosen to be 1000 ms,
100 ms and 10 ms (the latter being a rather extreme case for typical
non-recycled pulsars, which would be expected to be hosted in a binary
with an IMBH). For each of the parameters, we have
produced eight time series, spanning four different values of the
longitude of the periastron of the elliptical orbit (0, 90, 180 and
270 deg) and two values of the orbital phase (quadrature or
conjunction) for simulated observation.

After creating this large set of time series, we first applied
the standard search
(which maximizes the sensitivity for detecting isolated pulsars)
 and, later, the ``stack search'', which involves splitting each time
series into 16 segments, performing an FFT on each of them.
Then the resulting spectra
are summed applying a constant shift between frequency bins of adiacent
spectra, the shift corresponding to the change in the apparent spin
frequency between one segment of observation and the next (Faulkner et al.
2004). Many different bin shifts are explored in order to find the ones
which maximize the spectral signal-to-noise of a given periodicity in the
whole time series. At the sacrifice
of a certain loss of sensitivity (due to the incoherent sum of
spectra) this algorithm considerably improves the possibility of detecting
Doppler distorted signals at a reasonable computational cost.

For each simulated time series, we finally compared the
signal-to-noise ratio (s/n)$_{\rm obs}$ at which the simulated
pulsation period has been recovered by the search analysis with the
signal-to-noise ratio (s/n)$_{\rm sol}$ of the same signal if it were
been emitted by a solitary pulsar. For spin period $P>100$ ms, it
results that $\eta=(\rm {s/n})_{\rm obs}/(\rm {s/n})_{\rm sol}>0.8$ for all
cases, but those involving $e\simgreat 0.7$ and the closest orbital
separations. The application of the stack search allows to mantain a
relatively good sensitivity ($\eta \simgreat 0.6$) at most orbital
phases and eccentricities even for a 10 ms pulsars. In summary, only the
combination of very extreme eccentricities $e>0.9$, unfavorable orbital
phases (longitude of periastron $\sim 180$ deg and an observation
performed at about the epoch of periastron) and
orbital separations in the lower end of the assumed range decrease
the sensitivity of the search below $\eta=0.5.$ 

For a disc structure with thickness much smaller than its radial
extension, $\eta$ roughly represents the ratio between {the
galactic volumes explored to search} 
for the two types of sources; thus, assuming
a disc-like and uniform distribution for both pulsars,
i.e., those in IMBH-PSR
systems and the isolated pulsars, $\eta$ indicates approximately the relative
detection efficiency. In summary, given the orbital parameters of the
putative pulsars orbiting IMBHs, under the conditions
explored, we conclude that their detectability
by the Parkes Multibeam Pulsar survey has only been modestly affected
by their orbital motion. Other biases related to large-scale pulsar
surveys (discussed in \S 5.2) play the major role in selecting the
observed population from the intrinsic population of IMBH-PSR systems.

\section {Discussion} 

\subsection {The formation of an IMBH-PSR and its X-ray luminosity} 

The formation of an IMBH-PSR binary system relies mainly on four
working hypothesis: that an IMBH (i) forms in a young core-collapsed
dense star cluster, (ii) acquires, preferentially, a massive companion
star (with $M>15\msun$), given the dense environment in which it is
expected to form and live, (iii) experiences a phase of wind fed or
RLOF accretion that terminates when the donor star ends its life, and
that (iv) the donor star is sufficiently massive to explode leaving
behind a radio pulsar, as a remnant.

Hypothesis (iii) implies that an X-ray active phase preceeds the
formation of the IMBH-PSR binary.  Thus, IMBH-PSR progenitors can be
identified as X-ray sources.  The X-ray active phase is found to be
characterized by luminosities that spread over a broad range,
depending on the mass of the companion star and the initial orbital
separations. Two possibilities may occur: 
\noindent (a) 
If the star on the ZAMS is underfilling its Roche lobe, WFA
luminosities in excess of $L_{\rm {ULX}}\sim 3\times 10^{39}\ergs$ are
found only in presence of donors as massive as 30$\msun.$ With lighter
stars, WFA luminosities fall typically in the range between
$10^{37}-10^{39}\ergs,$ with the highest values coming from transfer
rates magnified by the focussing action of gravitational field of the
heavy IMBH on wind particles.  At separations upto several AU,
typical WFA luminosities fall short below $10^{38}\ergs$ so that,
based on luminosity arguments, it would be impossible to distinguish
an accreting IMBH from a stellar-mass black hole, under WFA.  WFA can
last as long as the entire time spent by the star on the main
sequence, $\tau_{\rm MS}\sim 1-10$ Myrs.  Expansion of the massive
star during nuclear evolution leads inevitably to a phase where the
star fills its Roche lobe. This occurs when the binary separation
$a\simless 10$ AU at the time the star is transiting the Hertzsprung gap
and/or ascending the giant branch.  At this moment, one expects a
dramatic increase of the mass transfer rate 
(see the spikes in Fig. 3 and 4, and also Portegies
Zwart et al. 2004b).  This is a short lived phase of a few $10^4-10^5$
yrs that accounts only for $\sim 5\%$ of $\tau_{\rm {MS}}$, in which
the binary can emit luminosities in excess of $10^{40}\ergs$,
characteristic of those bright ULXs in which we can be confident to
find an IMBH. Thus, the lifetime of a ULX, progenitor of an IMBH-PSR
system, $\tau_{\rm ULX}$ can be as
short as a few $10^4$ yrs.

\noindent (b) 
If, instead, the massive star overfills its Roche lobe near the zero
age, contact is maintained stably in the binary, and mean transfer
rates as large as $10^{-7}-10^{-5}\msun$ yr$^{-1}$ are found.  These
rates are large enough to guarantee the stability of the disc against
thermal-ionization perturbations. For donors more massive than $\sim
15\msun$ the mass transfer rate is higher than the Dubus' critical
value so that we argue in favor of persistent emission in binaries
with an IMBH and massive donors. The accretion luminosities inferred
at an averaged efficiency of 0.1, all exceed $10^{39}\ergs,$ but
only stars more massive than 20$\msun$ can guarantee luminosities
above $10^{40}\ergs$ during the main sequence. RLOF with massive donors ($>20\msun$) at the
ULX level lasts for a time $\tau_{\rm ULX}$ of few million years,
while lighter donors are found to brighten at the level of a few
$10^{39}\ergs$ for a time of $\simgreat 10$ Myrs.

Concerning the ultimate fate of the donor star we have seen that RLOF
on the ZAMS causes significant mass losses, so, one can estimate
that only stars heavier than 15$\msun$ and lighter than $30\msun$ may
end their lives as neutron stars and thus turn on as radio pulsars.
Lighter donors end probably their life as white dwarfs. These can still
produce ULX activity under transient conditions.

With hypothesis (iv), the formation of an IMBH-PSR system requires
that the supernova explosion of the companion to the IMBH does
not end with the disruption of the binary. Symmetric mass loss would
not unbind the system given the large inertia of the IMBH, but the
occurrence of an asymmetric natal kick can place the neutron star in
an unbound orbit.  Considering the pre-explosion orbital parameters
from our binary evolution models, we find that the probability of
survival is very high, around $99\%.$ The IMBH-PSR binary that
survives disruption would have preferential separations  between
1 and 10 AU, and a mild eccentricity.

\subsection {Perspectives of detection of the radiopulsar} 

In our study we have shown that binaries hosting an IMBH and a donor
star, massive enough to leave a pulsar as a relic, likely experience an
X-ray phase characterized by accretion luminosities ranging from well
below the Eddington luminosity of a one solar mass star, up to values
characterizing the bright tail of the ULX window. The life time
$\tau_{\rm ULX}$ over which a ULX phase is observed falls in the range
between 0.05 to 10 Myrs whereas $\tau_{\rm X}\simeq 10$ Myrs is the
characteristic lifetime of an accreting IMBH as ``normal'' X-ray
source.  While the formation of a binary IMBH-PSR system does not
necessarely require a transit through a ULX phase, a ULX phase can
inversly highlight a system that will evolve into an IMBH-PSR binary.
From the knowledge of the distribution function $P(a)$ of the initial
orbital separations we can calculate the ratio $\Re>1$ between the
number of binaries hosting an IMBH-massive star system ending as a
IMBH-PSR, to the number of binaries $N^{\rm PSR}_{\rm ULX}$ which
experience a bright ULX phase before becoming an IMBH-PSR. $\Re$ is:
\be
\Re=\frac{\int^{a_{\rm max}}_{r_{\rm t}} P(a)da}
         {\int^{a_{\rm ULX}}_{r_{\rm t}} P(a)da}
\ee
where $r_{\rm t}$ is the tidal distruption radius of the IMBH, $a_{\rm
max}$ is the orbital separation at which an X-ray phase (of arbitrary
low intensity) occurs ending with the formation of a pulsar bound to
the IMBH, and $a_{\rm ULX}$ is the initial orbital separation at which
a ULX phase sets in during binary evolution. If we suppose $P(a)$
is uniform, and take $a_{\rm max}\simeq 200$ AU as the limiting distance
above which a natal kick would unbind the system (for a chance
probability of $50\%$), and $a_{\rm ULX}\simeq 1$AU (the distance for
a RLOF phase during the main sequence), then we obtain an approximate estimate of $\Re\sim
200.$ This shows that the IMBH-PSR binaries whose progenitor
experienced a ULX phase may well represent only the tip of the iceberg
of the whole population of IMBH-PSR binaries. However, we note that the
estimate of $\Re$ is strongly affected by the uncertainity of the
maximum orbital separation allowed by an exchange interaction (which
corresponds to the cutoff of the distribution $P(a)$), and also by the
real form of the distribution $P(a)$. Detailed stellar dynamical
simulations are required to improve the knowledge of the crucial
factor $\Re$.

Provided $\Re$ is known, we can estimate the number of IMBH-PSR
binaries, from the simple formula $(\tau_{\rm PSR}/\tau_{\rm
X}) \Re f N_{\rm ULX}$ where $\tau_{\rm PSR}\sim 10$ Myr is the
lifetime of a radiopulsar, $N_{\rm ULX}$ the observed number of bright
ULX sources and $f\le 1$ is a fudge factor to compute $N_{\rm ULX}^{\rm
PSR}=fN_{\rm ULX}$ from the known number of ULXs. In order to calculate 
the number of potentially observable radiopulsars
$N_{\rm IMBH-PSR}$ orbiting an IMBH, one must also account for the strong
anisotropy in pulsar emission. Introducing the pulsar beaming factor
$b\sim 10$ (Lorimer 2001), and scaling the others quantities to the
reference values, it turns out
\be 
N_{\rm IMBH-PSR}\simeq \! 20 f \!
\left({\tau_{\rm PSR}\over {10\,\rm {Myrs}} }\right) 
\left( {10\,\rm {Myrs} \over \tau_{\rm X}}\right)
{\Re \over 200} {10\over b} N_{\rm ULX}
\label{imbh-psr}
\ee 
In section \S 4 we have shown that Doppler distortion of the
radio signal should not have hampered the detection of a putative
IMBH-PSR system by the Parkes Multibeam survey of the Galactic Disc
(Manchester et al. 2001).  However, other selection effects influence 
the detectability of a given radio pulsar during a survey, the most
relevant being the pulsar intrinsic luminosity and the survey sensitivity 
limit for a source at the given position in the Galaxy, in turn depending
 on the pulsar spin period, duty cycle, dispersion measure, and amount
 of scattering smearing. A population synthesis analysis combined with a
simulation, for the given survey, is necessary (e.g. Lorimer et
al. 1993) for assessing the ratio $\chi\le 1$ between the number of
expected detections and the number of potentially
observable radio pulsars (i.e. those having the radio beam sweeping the
line of sight to the Earth). Putative pulsar companions to IMBH and
typical isolated pulsars are populations of non-recycled pulsars
born and still residing in the Galactic disc. Assuming, in first
approximation, that the two populations both share the galactic
distribution and the same intrinsic properties (such as the
luminosity function), we can apply to the pulsars orbiting an IMBH the
``average'' value of $\chi$ properly calculated for the Parkes
Multibeam Survey (Manchester et al. 2001).  In particular, $\chi$ is in
the range $0.050-0.075$ for pulsars with luminosity greater than 1 mJy
kpc$^2$ (Vranesevic et al. 2004), and is $\sim 3$ times smaller, when 
including fainter sources having minimum luminosity $\sim 0.2$ mJy
kpc$^2.$

Correcting for this factor, and adopting the reference values of
equation (\ref{imbh-psr}), we expect that the Parkes Multibeam survey
should have detected
\be 
N_{\rm obs} = \chi N_{\rm IMBH-PSR} \simeq f N_{\rm ULX}
\label{psr-obs}
\ee 
\noindent 
pulsars orbiting an IMBH.  This estimate implies that the detection of
a radio pulsar orbiting an IMBH in the Galaxy is subjected to the same
small number statistics of the ULX population (whose detection in the
X-ray band is free from biases), even for the most favorable case
$f=1.$ In particular it is compatible with the current lack of
observed pulsars of this type. More sensitive surveys of the Galactic
disc (such as the ongoing P-Alpha survey at Arecibo) or targeted
surveys (searching deeply  for pulsars in the directions of young star
clusters) may results in a improved value of $\chi$, enhancing the
possibility of unveiling a pulsar orbiting an IMBH within few years. 
However another possibility is that the number of ULXs hosting an IMBH
is not so high as supposed, since as previously discussed, these
sources may be explained by simple stellar mass black holes, leaving
only a few number of ULXs with a true IMBH.
In this case we must rescale the value of the equation (16) 
to the unknown number $N^{'}_{ULX}<N_{ULX}$ of ULXs with a real IMBH,
diminishing our chance of a successfull detection.
Next generation radiotelescopes will be able to go well beyond the
Magellanic Clouds and eventually reach all the galaxies in the Local
Group (e.g. M31 with LOFAR, R\"ottgering 2003) or nearby galaxies
(e.g. M82 with SKA, Cordes et al. 2004). The much larger volume of
space explored will enclose some known bright 
ULX\footnote{The nearest known bright ULXs are at a distance of
  $\simeq 2-3$Mpc, whereas the ULX reported in the spiral
galaxy M33 seems to be an ordinary X-ray binary in a high state
rather than a true ULX (Foschini et al. 2004).}. If these future
instruments will be able to sample a sizeable fraction of the total
pulsar population in nearby external galaxies, they will greatly
increase the chance of discovering objects belonging to the very intriguing
class of binaries presented in this paper.

\vskip 1.0 cm
{\footnotesize{We would like to thank Onno Pols and Jasinta Dewi for the kind
support in the use of the Eggleton code, and Simon Portegies Zwart
for useful discussion on several aspect of this work. AP contribution
has been partially supported by the Italian Ministry for Education,
University and Research (MIUR) under grant PRIN-2004023189.
}}

\newcommand{\nat}{Nat}
\newcommand{\mnras}{MNRAS}
\newcommand{\aj}{AJ}
\newcommand{\pasp}{PASP}
\newcommand{\aap}{A\&A}
\newcommand{\apjl}{ApJ}

\end{document}